\documentclass[conference]{IEEEtran}
\IEEEoverridecommandlockouts
\usepackage{amsmath,amssymb,amsfonts}
\usepackage{algorithmic}
\usepackage{graphicx}
\usepackage{textcomp}
\usepackage[table,xcdraw]{xcolor}
\usepackage{balance}
\usepackage{soul}
\usepackage{booktabs}
\usepackage{multirow}
\usepackage{float}
\usepackage{url}
\usepackage{tcolorbox}
\usepackage{tabularx} 
\usepackage[noadjust]{cite}
\usepackage[hidelinks,bookmarks=true,pagebackref=true]{hyperref}
\usepackage{xurl}
\usepackage{enumitem}
\setlist[itemize,enumerate]{noitemsep, topsep=0pt, leftmargin=1.0em}
\usepackage[utf8]{inputenc}
\usepackage{newunicodechar,graphicx}
\DeclareRobustCommand{\okina}{%
  \raisebox{\dimexpr\fontcharht\font`A-\height}{%
    \scalebox{0.8}{`}%
  }%
}
\newunicodechar{ʻ}{\okina}

\usepackage{listings} % For code formatting
\usepackage{xcolor}   % For custom colors

\lstdefinestyle{pythonstyle}{
  language=Python,
  basicstyle=\ttfamily\small,    % Code font style and size
  keywordstyle=\bfseries\color{blue},   % Keywords style
  commentstyle=\itshape\color{green!50!black}, % Comment style
  stringstyle=\color{red},   % String style
  frame=single,               % Adds a frame around the code
  numbers=none,              % Line numbers on the left
  numberstyle=\tiny\color{gray}, % Line numbers style
  stepnumber=1,              % Number every line
  tabsize=4,                 % Tab size
  showstringspaces=false,    % Don't show spaces in strings
  breaklines=true,           % Automatic line breaking
  showspaces=false,
  captionpos=b,
  rulecolor=\color{black}
}
\def\thesubsubsectiondis{\unskip\arabic{subsubsection})}

\newcommand{\RQA}{\textbf{RQ1}: What are the prevailing patterns in case conventions and name length of method names within Jupyter Notebooks?}

\newcommand{\RQB}{\textbf{RQ2}:  What are the predominant grammatical patterns in method names within Jupyter Notebooks?}%, and how do these patterns compare to those found in method names across other programming languages?}

\newcommand{\RQC}{\textbf{RQ3}: What are the patterns and prevalence of abbreviations and acronyms in method names within Jupyter Notebooks?}%, and how do these compare to their usage in traditional programming contexts?}

\begin{document}

\title{\huge Method Names in Jupyter Notebooks: An Exploratory Study}

\newcommand\Mark[1]{\textsuperscript#1}

\author{
\IEEEauthorblockN{Carol Wong\Mark{1}, Gunnar Larsen\Mark{1}, Rocky Huang\Mark{1}, Bonita Sharif\Mark{2}, Anthony Peruma\Mark{1}}
\IEEEauthorblockA{
\Mark{1}University of Hawaiʻi at Mānoa, Honolulu, Hawaiʻi, USA \\
\Mark{2}University of Nebraska - Lincoln, Lincoln, Nebraska, USA \\
carolw8@hawaii.edu, gunnarrl@hawaii.edu, rhuang8@hawaii.edu, bsharif@unl.edu, peruma@hawaii.edu
}
}

\maketitle

\begin{abstract}
Method names play an important role in communicating the purpose and behavior of their functionality. Research has shown that high-quality names significantly improve code comprehension and the overall maintainability of software. However, these studies primarily focus on naming practices in traditional software development. There is limited research on naming patterns in Jupyter Notebooks, a popular environment for scientific computing and data analysis. In this exploratory study, we analyze the naming practices found in 691 methods across 384 Jupyter Notebooks, focusing on three key aspects: naming style conventions, grammatical composition, and the use of abbreviations and acronyms. Our findings reveal distinct characteristics of notebook method names, including a preference for conciseness and deviations from traditional naming patterns. We identified 68 unique grammatical patterns, with only 55.57\% of methods beginning with a verb. Further analysis revealed that half of the methods with return statements do not start with a verb. We also found that 30.39\% of method names contain abbreviations or acronyms, representing mathematical or statistical terms and image processing concepts, among others. We envision our findings contributing to developing specialized tools and techniques for evaluating and recommending high-quality names in scientific code and creating educational resources tailored to the notebook development community.
\end{abstract}

\section{Introduction}
\label{Section:Introduction}
Methods are fundamental elements of source code, encapsulating behavior and promoting code reuse and modularity \cite{mcconnell2004code}. They represent the smallest unit of functionality, enabling practitioners to break down complex processes into manageable, logical segments. This decomposition makes it easier to read, understand, and test the program \cite{martin2009clean,osherove2013art}. Equally important to having well-defined methods is the necessity for having method names that effectively communicate the purpose and functionality of a method \cite{Host2009,Host2007}. This is particularly important as developers spend 58-70\% of their time trying to comprehend code and only 5\% editing it \cite{Feitelson2023}.

Extensive research has examined identifier names, exploring their structure, semantics, evolution, and developers' perceptions \cite{LawrieICPC2006,ButlerCSMR2010,Li2020Renamings,Alsuhaibani2021,gresta2023naming, binkley-emse13}. These studies show that effective identifier names are crucial for improving code readability and maintainability and can improve code comprehension by as much as 19\% \cite{Hofmeister2017}. However, most of this research has primarily focused on code written in traditional programming environments by developers with formal computing backgrounds. As a result, there is a gap in our understanding of naming conventions in alternative environments like Jupyter Notebooks.

Unlike traditional Python programs, which are typically linear text files often spread across multiple files, Jupyter Notebooks are interactive documents divided into executable code cells interspersed with documentation and output cells. These 
notebooks offer an environment that supports interactive and iterative data exploration, prototyping, scratch work, creating tutorials and teaching aids, all within a single document \cite{Rule2018,Perkel2018}. This unique combination of factors can influence how practitioners, many of whom lack a formal background in computer science \cite{Kery2018,Santana2024}, name and structure their methods, particularly in the context of scientific programming \cite{Beg2021,Samuel2024}. 

While several studies have examined code quality in Jupyter Notebooks, focusing on style violations, reproducibility issues, and structural complexity and organization \cite{Wang2020,PimentelMSR2019,grotov2022large,Adams2023,Rule2018,Kery2018}, the quality of identifier names, a key factor in code comprehension, remains underexplored in notebook environments.

\subsection{Motivating Example}
Consider the methods shown in Listings \ref{lst:intro_code01} and \ref{lst:intro_code02}, which were extracted from a dataset of open-source Jupyter Notebooks \cite{grotov2022large}. In Listing \ref{lst:intro_code01}, the name is composed of a single term which happens to be an acronym, `rmse.' This stands for `Root Mean Square Error', which is a domain-specific term often used in fields like statistics, and machine learning and is generally unfamiliar to those outside of these fields. Using acronyms like `rmse' as method names can lead to confusion as it assumes that the users of the code will have prior knowledge of the term, which might not always be the case, especially for novices or interdisciplinary teams \cite{Hofmeister2017}. Further, the method name deviates from naming practices by not following a verb phrase pattern, which can lead to confusion regarding its action or behavior \cite{Alsuhaibani2021}. A more descriptive name like `compute\_root\_mean\_square\_error' would better convey the method's purpose. Listing \ref{lst:intro_code02},  in contrast, includes a verb (`search'), which ideally should be at the start of the name. The name has two other issues: (1) it does not follow Python's recommended snake\_case convention, and (2) it lacks a conventional boolean prefix, despite returning a boolean value. A more suitable name would be: `is\_game\_category\_present.'

\begin{lstlisting}[style=pythonstyle, caption={Method name lacking a verb prefix and composed of an acronym.}, label={lst:intro_code01}]
def rmse(predicted, actual):
    # formula for rmse
    residual = predicted - actual
    residual_sq = residual ** 2
    mean_sq = np.mean(residual_sq)
    rmse_value = np.sqrt(mean_sq)
    # return rmse_value
    return rmse_value
\end{lstlisting}

\begin{lstlisting}[style=pythonstyle, caption={Method name violating Python's snake\_case convention and not reflecting its actual behaviour.}, label={lst:intro_code02}]
def gamesearch(x):
 gamecategory=['computer games', 'video games', 'video-games', 'computer-games', 'mmorpg', 'nintendo', 'playstation']
 regex = re.compile("|".join(word for word in gamecategory), re.IGNORECASE)
 if regex.search(x['summary'] or x['title'] ):
    return 1
 else:
    return 0
\end{lstlisting}

These examples highlight the need to explore method naming practices in notebook environments, as they may not only share common naming violations with traditional programming environments (\cite{Host2009,Butler2009,Arnaoudova2016}) but also exhibit unique or more frequent violations due to their unique characteristics.

\subsection{Goal \& Research Questions}
\label{Section:Goal}
The paper presents an exploratory study to enhance our understanding of naming practices in scientific programs, specifically focusing on code contained within Jupyter Notebooks. \textit{Our goal is to examine the characteristics of method names in notebook environments, including their grammatical structure, case conventions, name length, and use of abbreviations and acronyms}. We focus our evaluation on method names as they encapsulate behavior and represent primary units of functionality within code \cite{Host2009}. Further, method names play a crucial role in conveying the functionality and intent of code segments, directly impacting program comprehension and maintainability \cite{Allamanis2015, Host2007}. By understanding how practitioners typically craft method names in scientific programs, we envision our findings leading to: (1) the development of tools and techniques in method name recommendation and appraisal in the context of scientific programming and (2) the development of educational materials to train scientific programmers in effective naming practices. Furthermore, our focus on naming patterns provides the groundwork for exploring broader issues such as notebook reproducibility and collaboration practices.

\vspace{1.3mm}
\noindent We aim to answer the following research questions (RQs): 

\vspace{1.3mm}
\noindent\textbf{\RQA} Case conventions and identifier name length are essential to code readability and understandability \cite{Hofmeister2017,Binkley2009, binkley-emse13}. This RQ examines how practitioners balance descriptiveness and conciseness in their method names.

\vspace{1.3mm}
\noindent\textbf{\RQB} In this RQ, we aim to identify how practitioners construct method names, particularly their grammatical structure, in notebook environments. Instead of analyzing individual terms in the name, we examine part-of-speech patterns, which comprise a finite set. This approach allows us to uncover common naming conventions, as the same part-of-speech pattern can represent numerous semantically distinct yet syntactically similar names. 

\vspace{1.3mm}
\noindent\textbf{\RQC} This RQ explores the prevalence and types of abbreviations and acronyms practitioners use when crafting method names in notebook environments. Since notebooks are frequently employed for scientific and domain-specific programming \cite{Samuel2024,Beg2021}, understanding these naming patterns can reveal how practitioners balance naming conciseness with readability in specialized contexts.

\subsection{Contributions}
The main contributions of this work are:
\begin{itemize}
    \item A detailed analysis of the grammatical structure of method names in Jupyter Notebooks, complemented by a manually annotated dataset of part-of-speech tags in these identifiers.
    \item An improved understanding of the use of abbreviations and acronyms in notebook method names and their role in conveying information.
    \item Our findings provide a strong initial step towards enhancing the readability and understandability of scientific software by identifying opportunities for future research, including the development of automated tools tailored to notebook environments and scientific code analysis.
\end{itemize}

\section{Related Work}
\label{Section:related}

\subsection{Code Quality in Jupyter Notebooks}
In their analysis of 1,982 Python-based Jupyter Notebooks, Wang et al. \cite{Wang2020} report that the notebooks have a high error ratio (36.26\%) against PEP8 guidelines compared to independent Python scripts (13.40\%). The authors also highlight the prevalence of unused variables and the use of deprecated functions. Grotov et al. \cite{grotov2022large} conduct a large-scale comparison of Python code in Jupyter Notebooks and traditional scripts, focusing on structural and stylistic differences. Their findings indicate that notebooks contain 1.4 times more stylistic issues and tend to have more entangled, though structurally simpler, code. Additionally, the authors report that notebooks contain fewer unique user-defined functions but use them more frequently. In comparing machine learning code quality between Python scripts and Jupyter Notebooks, Adams et al. \cite{Adams2023} find that while Jupyter notebooks are larger and less complex, they exhibit higher coupling and more frequent use of built-in functions, and contrast with Grotov et al. \cite{grotov2022large} showing that scripts have more style issues per file.

In a study of 1,159,166 unique Jupyter Notebooks, Pimentel et al. \cite{PimentelMSR2019} report that most display good practices like using markdown cells, visualizations, and meaningful filenames. However, issues such as out-of-order code cells and non-executable cells hinder reproducibility. %Patra and Pradel \cite{Patra2022} highlight the importance of meaningful names in dynamically typed languages through their technique, Nalin, which automatically detects name-value inconsistencies. %Nalin combines dynamic analysis with neural machine learning to identify mismatches between variable names and their runtime values. 
A study involving a multivocal literature review, interviews with professional data scientists, and an analysis of 1,380 Jupyter Notebooks from Kaggle by Quaranta et al. \cite{Quaranta2022} identifies 17 best practices across six themes for collaboration with computational notebooks in data science. Among these best practices, the authors highlight the importance of giving notebooks meaningful names and adhering to widely accepted coding conventions, such as PEP 8 for Python. In their examination of computational notebooks, Rule et al. \cite{Rule2018} highlight that many notebooks are viewed as personal, exploratory, messy, and lacking explanatory text. Due to these factors, notebooks often need to be cleaned and annotated before sharing, which is perceived as tedious and time-consuming.

\subsection{Grammar Patterns \& Abbreviation and Acronyms}
Newman et al. \cite{NewmanCoRR2020} examine 1,335 identifiers from 20 Java and C/C++ open-source systems to identify common grammar patterns used in identifier names. The authors find that a frequently observed grammar pattern is the verb phrase, primarily used in function names or boolean variables. 
A survey of 1,162 professional developers by Alsuhaibani et al. \cite{Alsuhaibani2021} reports that most developers agree that method names consisting of multiple words should be presented in a grammatically correct structure and that every method name needs to contain at least one verb to enhance understandability. 
In a study of JUnit-based unit test methods, Peruma et al. \cite{PerumaICPC2021} identify the most prevalent grammar patterns, revealing that the top five most common patterns for test method names are all verb phrase patterns. In their study of identifier renaming in Java projects, Arnaoudova et al.\cite{Arnaoudova2014} reports that 76\% of the classified renamings did not involve a part of speech change and 13\% involved singular/plural changes. The majority (83\%) involved other parts of speech changes, such as from nouns to adjectives. Butler et al. \cite{Butler2011} analyze 120,000 identifier names by identifying common and project-specific naming conventions. Their work focuses primarily on class names and finds that a majority are composed of noun phrases. Caprile and Tonella \cite{Caprile1999} analyze one industrial and nine public domain applications, performing a lexical, syntactic, and semantic analysis of function identifier names. They develop a regular grammar that effectively parses many identifiers from their database of 3,304 identifiers. Host et al. \cite{Host2009PhraseBook} created a phrase book to assist Java programmers in generating meaningful identifier names, highlighting the importance of longer, descriptive names for clarity on a method's behavior.

In their analysis of abbreviation expansion, Newman et al. \cite{NewmanICSME2019} examined 861 abbreviation-expansion pairs and found that 19\% of words in multi-word expansions are non-adjacent. Non-adjacency in expansions leads to ambiguity in the meaning of an abbreviation.
In a study where participants locate semantic errors in code, Hofmeister et al. \cite{Hofmeister2017} find that code with identifier names consisting of full words is comprehended 19\% faster than code with identifier names that consist of abbreviations and single letters. The authors also discover that abbreviations and single letters may signal lower-quality code. In a survey with 128 participants, including students and professionals, Lawrie et al. \cite{LawrieICPC2006} find that full-word identifiers are more comprehensible than single-letter identifiers.

\subsection{Summary}
Prior studies on identifier naming, especially regarding grammar patterns, abbreviations, and acronyms, have primarily concentrated on traditional software development. However, there is a limited understanding of these practices within Jupyter Notebooks. Existing research on notebooks has mainly focused on structural aspects, such as PEP8 violations and reproducibility issues, rather than on naming conventions, rather than naming practices. This study aims to bridge these areas by providing the first systematic analysis of method naming practices in notebook environments, examining naming styles, grammatical patterns, and the use of abbreviations and acronyms. Understanding these patterns will help in developing educational resources and specialized tools to improve code quality and the overall maintainability of notebooks.

\section{Study Design}
\label{Section:method}
This section provides details on the design of our study.

\subsection{Source Dataset}
The study utilizes an existing dataset of 847,881 permissively licensed Jupyter notebooks with Python as the specified language. The notebooks were collected from GitHub and made available by Grotov et al. \cite{grotov2022large} for their study on analyzing Python code within Jupyter Notebooks. The dataset is available as a PostgreSQL database, where each markdown and code cell is represented as rows in tables. This dataset has also been used in prior research studies \cite{Montandon2023,Grotov2023}.

\subsection{Notebook Extraction} 
While a manual analysis of the entire dataset of 847,881 Jupyter notebooks is comprehensive, it can be impractical. Hence, for this study, we extracted a statistically significant number of random notebooks from the dataset. Using the \texttt{random()} function in PostgreSQL, we extracted a statistically significant random sample from the source dataset with a 95\% confidence level and 5\% margin of error, resulting in 384 notebooks. A custom Python script was created to extract and reconstruct these 384 notebooks from the database rows.

\subsection{Method Name Extraction} 
We built a custom Python script to extract method details from the set of 384 notebooks. Our code first converted the notebook to a Python script and used Python's \texttt{ast} module to parse through the notebooks, collecting the method names, parameters, and return statements. Since this is an exploratory study, manually analyzing all methods across the 384 notebooks was unfeasible. Therefore, we selected two methods from each notebook for analysis. This sampling strategy enables us to examine a diverse range of notebooks while keeping the analysis workload manageable and capturing potential variations in naming practices. If a notebook contained only one method, that method was chosen for analysis. In total, we extracted 695 method names for manual review.

\subsection{Manual Analysis}
As this is an exploratory study, we choose to conduct a manual analysis of method names instead of relying on an automated approach. The manual approach allows for deeper insights into the semantic and syntactic nuances of these names. Furthermore, current part-of-speech taggers still need considerable improvements to effectively analyze identifiers \cite{Olney2016,NewmanCoRR2020,Newman2022}. Likewise, abbreviation expansion tools/techniques often face limitations in their expansion accuracy \cite{NewmanICSME2019}.

We manually analyze each of the extracted 695 method names to answer the three RQs posed in Section \ref{Section:Goal}. As part of this analysis, for each of the 695 method names, we review the method body and surrounding code. Additionally, we also review the entire notebook if the surrounding code does not provide sufficient context. In the process, we identified four names that were non-English and not recognized as domain or technology terms, which we excluded. After this exclusion, we were left with 691 method names to answer our RQs. The manual analysis was carried out by four authors, each with over four years of programming and research experience in program comprehension. Below, we describe the different procedures used in the manual analysis.

\subsubsection{\textbf{Grammar Pattern Analysis}}
To determine the grammatical structure of the method names, we utilized the part-of-speech tags defined by Newman et al. \cite{NewmanCoRR2020}. These tags are specifically tailored for analyzing source code identifiers and include: noun (\textbf{N}), noun modifier (\textbf{NM}), verb (\textbf{V}), verb modifier (\textbf{VM}), preposition (\textbf{P}), determiner (\textbf{DT}), conjunction (CJ), pronoun (\textbf{PR}), digit (\textbf{D}), and preamble (\textbf{PRE}). Further, the \textbf{NM} tag is used for both adjectives and noun-adjuncts, which are nouns used as adjectives to modify other nouns. 

Using these tags, three of the authors individually annotated each method name and tagged each term with a part-of-speech label. Before annotating, the authors received training on the tags to ensure a consistent understanding of the part-of-speech classifications. This training included resources and examples (such as \cite{SCANLCatalogue}) and discussions about the appropriate application of each tag in the context of method names.   

Each annotator reviewed the notebook where the method name was located to gather context on the method by examining the method's body, as well as any surrounding code, comments, or markdown cells. Context is vital when determining the grammar pattern of a method name, as the same term may have a different part-of-speech tag depending on the context. For example, the term `test' could either be a verb in \texttt{testModel} where the method evaluates the model, or a noun modifier in \texttt{testData} where it specifies the test dataset. To ensure the reliability of results, the authors cross-validated their annotations to ensure consistency and resolved disagreements through discussion, which included reviewing the code. Additionally, we calculated the inter-annotator agreement on the original annotations using Fleiss' Kappa (\cite{fleiss1971measuring}), obtaining a value of 0.813, which shows a very strong agreement among the annotators \cite{kappaInterpret}.

\subsubsection{\textbf{Name Composition Analysis}}
As part of identifying the part-of-speech tags, the annotators also split each method name into its individual terms and annotated the case convention styling associated with the name, and the method's return type. Any conflicts in this process were resolved through discussion. We achieved an inter-annotator agreement value of 0.984 using Fleiss' Kappa for term splitting and 0.97 for return types.

\subsubsection{\textbf{Abbreviation and Acronym Analysis}}
Following a similar approach, the annotators also reviewed method names to determine the existence of abbreviations or acronyms in the method's name. Next, for each identified abbreviation/acronym, the annotators reviewed the surrounding code (including comments) and markdown cells, and if that was not sufficient, the entire notebook to identify the expansion of the abbreviation or acronym. Similar to the prior analysis activities, conflicts were resolved through discussion. Using Fleiss' Kappa, we calculated an inter-annotator agreement for the detection of abbreviations and acronyms at 0.966 and an agreement of 0.987 for their expansions.

\subsubsection*{\textbf{Dataset}}
The grammar patterns and abbreviations/acronyms dataset, with the associated code, is available at \cite{ArtifactPackage}.

\section{Results}
\label{Section:results}
In this section, we present our findings by answering our RQs. Due to space constraints, in some instances, we report on frequently occurring instances; the complete dataset is available in our artifact package at: \cite{ArtifactPackage}.

\subsection*{\RQA}
Our analysis of the case conventions used by practitioners when naming methods shows that the majority of names follow the snake\_case convention, with 477 instances, accounting for 69.03\% of the total.  Table \ref{Table:case} presents the distribution of the different case convention types in our dataset. As part of our annotation process, method names composed solely of lowercase characters without any underscores were classified under the lowercase category, while names consisting entirely of uppercase characters were categorized as screaming case.

In examining the terms making up a method's name in our dataset, we observe that 337 (or 48.77\%) of the names consist of two terms, followed by 159 (23.01\%) and 149 (21.56\%) of names consisting of one and three terms, respectively. The median number of terms in a name is two, while the maximum number of terms is six, which occurs twice in our dataset. On average, there are 5.22 characters per term in a name. There are nine methods composed of only a single character (after ignoring digits in the name). 

Our dataset contains 22 methods with digits in their names. Among these, the digit `2' is the most common, occurring in 13 instances. The most common position for a digit is the second term, such as in \texttt{top\_3\_accuracy}. The digit is often used to indicate sequential order or as a distinguisher (e.g., \texttt{q1} and \texttt{q2}). Additionally, it may represent machine learning-specific terms, such as \texttt{f2\_score\_thr}, where the `F2' score is a metric used to evaluate a model's performance) or serve as a shorthand for the preposition `to' (e.g., \texttt{Ang2Pix}).

Finally, when analyzing the first term in the names, we identified that the most frequent terms are `get,' `plot,' and `load,' with 63 (9.12\%), 35 (5.07\%), and 22 (3.18\%) instances, respectively. In total, there are 283 unique starting terms.

\begin{table}
\centering
\caption{Method name case convention types in our dataset.}\vspace{-2mm}
\label{Table:case}
\resizebox{\columnwidth}{!}{%
\begin{tabular}{@{}lrrl@{}}
\toprule
\multicolumn{1}{c}{\textbf{Case Convention}} & \multicolumn{1}{c}{\textbf{Count}} & \multicolumn{1}{c}{\textbf{Percentage}} & \multicolumn{1}{c}{\textbf{Example}} \\ \midrule
Snake Case     & 477 & 69.03\% & \texttt{pick\_top\_n}        \\ \midrule
Lowercase      & 146 & 21.13\% & \texttt{forward}             \\ \midrule
Camel Case     & 33  & 4.78\%  & \texttt{plotData} \\ \midrule
Pascal Case    & 12  & 1.74\%  & \texttt{HalfPlus}            \\ \midrule
Screaming Case & 17  & 2.46\%  & \texttt{SVD}                 \\ \midrule
Mixed Case     & 6   & 0.87\%  & \texttt{Read\_File}          \\ \bottomrule
\end{tabular}%
}\vspace{-3mm}
\end{table}

\begin{tcolorbox}[top=0.5pt,bottom=0.5pt,left=1pt,right=1pt]
\textbf{RQ1 Summary.}
Most practitioners adhere to Python's PEP 8 convention for method naming, using snake\_case for multi-term names and lowercase for single-term names. Typically, method names consist of two terms (48.77\%), with the median length also being two terms.
\end{tcolorbox}

\subsection*{\RQB}
In this RQ, we report on the grammatical patterns observed while manually analyzing method names. We begin by briefly discussing the volume of unique grammatical patterns, then focusing on the most frequently occurring patterns, and finally highlighting some notable patterns.

Our analysis identifies a total of 68 unique grammar patterns. Among these, 384 instances (55.57\%) begin with a verb (e.g., \texttt{compute\_series}), while 271 instances (39.22\%) begin with a noun or noun modifier (e.g., \texttt{triplet\_acc}). Additionally, 36 instances (5.21\%) start with another part-of-speech tag, such as a preposition (e.g., \texttt{on\_epoch\_end}). 

Given that verb-initial patterns form the majority and are widely recommended in method naming conventions, we conducted a deeper analysis of these patterns. We observe 23 unique patterns where the verb is immediately followed by either a noun or a sequence of noun modifiers leading to a noun or plural noun (i.e., V NM* (N$|$NPL)). These 23 patterns account for 325 instances (i.e., there are 325 method names that follow this naming pattern). For example, the method name \texttt{get\_year} follows a V N pattern, while \texttt{create\_data\_model} follows a V NM N pattern. Additionally, through rarely occurring, there are instances where the verb is followed by a determiner, preposition, and conjunction, each occurring only twice in the dataset.  

Additionally, we manually examined a subset of 248 methods to understand their typical usage in notebooks, utilizing a 95\% confidence level and a 5\% margin of error. The methods were categorized based on how their return values were utilized: 122 methods (49.19\%) had their return values assigned to a variable (e.g., `factor = get\_factor()`), while 32 methods (12.9\%) used return values in expressions (e.g., `sum = get\_total() + calc\_tax()`). There were five methods (2.02\%) that were used in both ways, and 55 methods (22.18\%) had no return values. Additionally, 88 methods (35.89\%) were either unused or called without saving their results.

\subsubsection*{\noindent\textbf{1) Frequent Grammar Patterns}}\hfill\\
Next, we report on the common grammar patterns that appear frequently in our dataset. Table \ref{Table:PrefixPatterns} presents the six most prevalent grammar patterns found in method names, which we discuss in detail below. The complete list of identified grammar patterns is available in our artifact package.

\begin{table*}
\centering
\caption{The top six frequently occurring grammar patterns for method names.} \vspace{-2mm}
\label{Table:PrefixPatterns}
\begin{tabular}{llrrl}
\toprule
\textbf{Grammar Pattern} & \textbf{Phrase Type}          & \textbf{Count} & \textbf{Percentage} & \textbf{Example}                                                                                                                                 \\\hline
V N             & Verb Phrase   & 135   & 19.54\%           & \begin{tabular}[c]{@{}l@{}}\texttt{process\_image}\\  \textit{`process' (vberb) an image (noun) by resizing, cropping, and normalizing it.}%`process' is a verb and `image' is a noun
\end{tabular}                                      \\ \midrule
N               & Noun Phrase   & 88    & 12.74\%           & \begin{tabular}[c]{@{}l@{}}\texttt{timer}\\ \textit{`timer' (noun) prints the execution time of code blocks.}% is a noun
\end{tabular}                                                                       \\ \midrule
NM N            & Noun Phrase   & 72    & 10.42\%          & \begin{tabular}[c]{@{}l@{}}\texttt{neural\_net}\\ \textit{the method creates and configures a neural network; `net' is a noun and `neural'} \\\textit{is an adjective describing the type of network.} \end{tabular} \\ \midrule
V NPL           & Verb Phrase   & 65    & 9.41\%            & \begin{tabular}[c]{@{}l@{}}\texttt{load\_images}\\ \textit{`load' (verb) `images' (plural noun) from a directory into a list.}\end{tabular}                                   \\ \midrule
V NM N          & Verb Phrase   & 64    & 9.26\%            & \begin{tabular}[c]{@{}l@{}}\texttt{parse\_csv\_row}\\ \textit{`parse' (verb) a `csv' (noun modifier) `row' (noun) into a dictionary of features.}\end{tabular}         \\ \midrule
V               & Verb Sequence & 48    & 6.94\%            & \begin{tabular}[c]{@{}l@{}}\texttt{normalize}\\ \textit{`normalize' (verb) data by scaling its values.} \end{tabular} \\ \bottomrule            \end{tabular}\vspace{-4mm}
\end{table*}

We first elaborate on the grammar patterns starting with a verb and then those starting with a noun.

\textbf{V N}, \textbf{V NPL}, and \textbf{V NM N} are known as verb phrases, where the verb is being applied to the noun that follows. In these patterns, the verb represents the action being performed, while the subsequent noun (with optional modifiers) represents the object receiving that action. 

In our dataset, \textbf{V N}  is the most common grammar pattern with 135 occurrences. An example of the grammar pattern is \texttt{process\_image}, where `process' is the verb of the method and `image' is the noun that the verb is being applied to. Specifically, the purpose of this method is to scale, crop, and normalize the provided image. Examining our dataset, we also observe other derivatives of this pattern, such as a prepositional phrase (\textbf{P NM* (N$|$NPL)}) with a leading verb. By adding the prepositional phrase, the names provide more context as to the purpose and result of the function. For example, consider the method \texttt{write\_df\_to\_s3}, which follows a V N P N pattern. Here, `write' is the action performed on the noun `df' (dataframe), and the prepositional phrase `to\_s3' indicates the action's destination. This context clarifies what the method does (writes a dataframe) and where the result is stored (S3).

The fourth and fifth most common patterns are \textbf{V NPL} and \textbf{V NM N} with 65 and 64 instances, respectively. \textbf{V NPL} is a verb phrase pattern where an action (i.e., verb) is being applied to multiple objects (i.e., the plural noun that follows). An examination of methods in our dataset following this grammar pattern shows that these methods are often used to retrieve information and would be expected to return an array or collection of items. For example, as per the markdown cell contents, the method \texttt{get\_batches} ``\textit{returns batches of input and target data for our model}.'' 

The grammar pattern \textbf{V NM N} is a verb phrase pattern with a verb being applied to a noun that has a noun modifier. The noun modifier adds specificity to the noun, making it clearer what type of object the noun is that this action is being applied to. One instance of this is \texttt{find\_true\_label}, where the term `find' is the verb, `true' is the noun modifier, and `label' is the noun. This grammar pattern allows for methods to communicate their function with more specificity while maintaining a clear action-object relationship. The inclusion of the noun modifier not only indicates the action being performed and the object involved but also specifies the particular type or variation of that object.

\textbf{V} is a verb sequence pattern, which is defined as one or more verbs with no noun phrase. In our dataset, we have 48 instances of method names that are composed of only a single verb. Typically, these names are generic and provide less semantic information about their specific functionality compared to verb-noun patterns. For example, the method \texttt{clean} indicates an action but does not specify what is being cleaned. It is only once we examine the method body that we are made aware that the method cleans textual data by removing stopwords, punctuations, and other related tasks. In contrast, our dataset also contains a method named \texttt{clean\_text}, which performs similar actions as \texttt{clean}, but has a more descriptive name that conveys both the action and its target object. While single-verb names are concise and emphasize the action being performed, they can also potentially lead to ambiguity and increased cognitive load, impeding code comprehension and maintenance activities \cite{Host2007,Binkley2011,Hofmeister2017}.

\textbf{N} is a noun phrase pattern where methods contain a single noun term, without any adjectives or noun modifiers. These method names are typically either generic or require domain knowledge to understand their functionality. In our dataset, we observed 88 instances (12.74\%) of this pattern, making it the second most common pattern overall. These names often fail to provide sufficient context to understand their function or return type without examining the implementation or documentation. For example, the method name \texttt{model}, is unclear on the type of model that this function is referring to or the action associated with the model. However, upon examination of the method body, it becomes clear that the method creates and returns a Keras model instance based on the inputs. Additionally, falling under this pattern are method names composed of a single character (e.g., \texttt{f}) and names that correspond to abbreviations (e.g., \texttt{mse} for mean squared error) and acronyms (e.g., \texttt{vol} for volume). The widespread use of this naming pattern in our dataset highlights a common yet problematic practice among practitioners, who often rely on implicit context or surrounding documentation rather than explicitly defining a method's purpose through its name. This ambiguity can hinder code comprehension and maintenance, especially in machine learning notebooks where various models and data structures may coexist.

\textbf{NM N} is a noun phrase pattern where a noun modifier is added to further describe the head noun. In our dataset, this pattern appears 72 times (10.42\%), making it the third most common pattern. This structure allows specialization of the noun and can provide more context, improving to the understandability of the method's purpose. An example of this is \texttt{author\_url} where `url' is the head noun and `author' is the descriptive modifier, indicating that the URL specifically applies to an author. However, while this pattern is more descriptive than single-noun patterns, it still lacks explicit action information. For instance, \texttt{author\_url} does not indicate the action performed on the author's URL. Examining the method body, we observe that the purpose of this method is to dynamically create and return a URL. The prevalence of this pattern in our dataset suggests that practitioners
value specificity in method names, even without explicit verbs. 
However, this pattern still relies on implicit context or documentation to convey the method's actual behavior.

\subsubsection*{\noindent\textbf{2) Other Notable Patterns}}\hfill\\
In our examination of the 68 unique patterns, we observed several patterns that, although less frequent, provide insights into method naming practices in Jupyter Notebooks.

\textbf{Unconventional Word Order.}
We observed 8 instances where the order of the grammar pattern in method names deviated from conventional syntax. These include 5 instances of \textbf{N NM} and 3 instances of \textbf{N NM NM}. For example, the method \texttt{reg\_non\_param} performs a non-parametric regression, and has ``Non-parametric regression'' noted within a markdown cell above the method. However, the method name follows an \textbf{N NM NM} pattern, placing the noun (`reg') before the noun modifiers (`non' and `param'). This is atypical, as common code convention recommends that noun modifiers should precede the noun to align with natural language order \cite{NewmanCoRR2020}, such as ``non\_param\_reg.'' Reversing the order has the potential to obscure the method's intended purpose, as readers expect modifiers to qualify the head noun that follows it.

\textbf{Verb Terminating Patterns.}
Similar to the above, another notable category is patterns that end with a verb (e.g., \textbf{N V}, \textbf{V V}), accounting for 32 instances in our dataset. From these, 20 instances fall under \textbf{N V}, while four instances fall under \textbf{V V} and three under \textbf{VM V}. For example, the method \texttt{prediction\_run} follows the \textbf{N V} pattern, where the noun precedes the action being performed on it, in contrast to the more common verb phrase pattern. When examining the implementation, we found that this method generates a sequence of predictions based on an RNN model. A name following the verb phrase pattern, such as ``run\_prediction'' or ``generate\_predictions'', would better align with standard naming practices while maintaining semantic clarity. An example of the \textbf{V V} pattern is \texttt{combine\_select}, where combined data is retrieved. This composition of two verbs has the potential to result in ambiguity about the order of operations. Lastly, \texttt{maybe\_download} is an example of the \textbf{VM V} pattern where the methods' purpose is to ``\textit{download a file if not present}.'' Here, the verb modifier `maybe' indicates the conditional nature of the download operation.

\textbf{Preposition Patterns.}
Another category is the patterns that contain prepositions. Prepositions often provide specificity into how the verb and noun phrases relate to each other. Among the 46 instances of patterns with prepositions, the most common term used is `to,' which appears 18 times. The preposition `to' is often used in method names that convert one object into another format or type. These typically begin with a verb or noun phrase followed by a prepositional phrase, such as \textbf{N P N} or \textbf{V N P N}. In these cases, the noun before the preposition indicates the starting format, and the noun following the preposition specifies the target type. For example, in the method \texttt{unicode\_to\_ascii}, the preposition `to' indicates that Unicode text is converted to ASCII text. The second-most common preposition is `from,' which accounts for 13 instances. This preposition is most associated with methods that obtain or extract information. An example of this usage is \texttt{get\_face\_embeddings\_from\_image}, where the preposition `from' notes the source (i.e., an image) from which the information (face embeddings) is being extracted. 

\textbf{Conjunction Patterns.}
Grammar patterns containing conjunctions are another notable pattern, where they are often used to link two phrases or terms together. There are a total of nine method names that contain conjunctions, with the most common conjunction being `and' appearing five times. In some cases, `and' is used to indicate that the methods perform two operations. For example, in the method \texttt{generate\_and\_save\_images}, the method generates and subsequently saves the generated images, denoting a sequence of operations. In other instances, conjunctions link terms of data that are returned. For instance, \texttt{AgeAndRHR} returns the age and resting heart rate of a subject.

\textbf{Method Returns.}
Our dataset contains 533 methods with return statements. From this, 267 (50.1\%) method names do not start with a verb. The two most common non-verb starting grammar patterns are \textbf{N} and \textbf{NM N}, with return statements appearing 75 and 58 times, respectively. These patterns often name methods after their return value rather than their action. For example, the method \texttt{accuracy} uses a single noun that represents its return value, calculates and returns the \textit{``\% of values that were correctly predicted}.'' Additionally, we noted another pattern related to method returns: method names that end with a verb but do not start with one. There are 20 instances of this pattern, such as \textbf{N V}, which occur 16 times. An example of this pattern is \texttt{gamesearch()}, which we discuss in Section \ref{Section:Introduction}.

\subsubsection*{\noindent\textbf{3) Ambiguous Parts-of-Speech}}\hfill\\
Ambiguity in the part-of-speech tag of a method name is common, especially in single-term method names. This particularly occurs with terms that can function as either nouns or verbs, often requiring examination of the code context and surrounding markdown cells to understand their intended grammatical role. For example, the method name \texttt{square} is vague in its function. The term `square' can either be a noun, referring to a geometric shape, or a verb, indicating the action of mathematically squaring a value. In this case, the method has the latter functionality, which we identified by examining the implementation. Similarly, the method name \texttt{fit} can either be a noun referring to a statistical fit of the data or a verb indicating that the method fits a model to the data. The comment within the method indicates that it is the latter, confirming its use as a verb. These polysemous names make it harder to understand the purpose of a method \cite{Host2011}.

\begin{tcolorbox}[top=0.5pt,bottom=0.5pt,left=1pt,right=1pt]
\textbf{RQ2 Summary.}
Despite representing a majority, only 55.57\% of method names adhere to the conventional practice of beginning with verbs. Meanwhile, 39.22\% start with nouns or noun modifiers, indicating widespread deviation from established naming practices. The most prevalent pattern was verb-noun (V N) at 19.54\%, followed by single nouns (N) at 12.74\%. The high frequency of non-verb-initial patterns, particularly in methods with return values (50.1\%), suggests a tendency to name methods based on their output rather than their action. Notable issues include ambiguous single-term names, reversed word orders, and unconventional patterns like methods ending with verbs, which can impact code comprehension.
\end{tcolorbox}

\subsection*{\RQC}
The goal of this RQ is to explore the occurrence of abbreviations and acronyms in method names and broadly understand the different types of categories that these terms can represent and the extent to which their expansions are readily available. 

In total, 210 (or 30.39\%)  method names include abbreviations or acronyms, with 19 of those names containing more than one abbreviation or acronym (e.g., \texttt{read\_gaia\_psf\_sdss}). From these 210 method names, we identified 230 abbreviations or acronyms. An analysis of these terms reveals the following nine high-level categories which they fall into:   
\begin{itemize}
    \item \textbf{Mathematics and Statistics} is the dominant category, comprising 67 abbreviations and acronyms. Within this category are error metrics such as mean squared error (`mse'), statistical measures like correlation (`corr'), and mathematical operations such as cosine (`cos').
    
    \item \textbf{Image Processing} terms occur 26 times, with the most common terms being abbreviations of image (`img' and `fig'). Other terms within this category include specialized image algorithms/techniques, such as point spread function (`psf') and histogram of gradients (`hog'). 
    
    \item \textbf{Programming}-related terms account for 21 abbreviations and acronyms. A notable pattern is using abbreviations and acronyms for the term function (e.g., `func,' `fn,' and `f') in the method name, such as in \texttt{model\_rnn\_fn}, creating redundancy in naming conventions.
    
    \item \textbf{Machine Learning and AI} terms occur 20 times and represent algorithms such as support vector machine (`svm') and k-nearest neighbors (`knn'), neural network architectures, such as recurrent neural network (`rnn'), and deep convolutional generative adversarial network (`dcgan'). Additionally, this category also includes terms related to training-related operations, such as learning rate (`lr').% and multi sample dropout (`msd'). 

    \item \textbf{Data Structure} terms account for  15 instances and represent structures, such as arrays (`arr'), dictionaries (`dict'), and dataframes (`df'). Also included are terms that represent both data structure and file types, such as extensible markup language (`xml') and comma separated value (`csv').

    \item \textbf{Database and Services} terms constitute 12 entries and represent external datasets such as the internet movie database (`imbd'), comparative toxicogenomics database (`ctd'), and storage services like Amazon Simple Storage Service (`s3').

    \item \textbf{Action Verbs} are an interesting category as they represent the use of abbreviations to describe the behavior or purpose of the method and typically appear at the beginning of the name. We encounter 14 instances in our dataset. For example, in the method name \texttt{'calc\_harmonics'}, the abbreviation `calc' expands to calculate. Other examples include `inc' (increase), `dec' (decrease),  and `gen' (generate).

    \item \textbf{Domain Specific} terms are terms unique to specific scientific or technical fields associated with the notebook they are found in, and they account for 14 occurrences in our dataset.  For example, `sdss' is an abbreviation for Sloan Digital Sky Survey, which is related to the domains of astronomy and astrophysics, while `pcr' (polymerase chain reaction) is part of the molecular biology field. 

    \item \textbf{Other} represents 41 abbreviations and acronyms and represents a collection of terms that do not fit into any of the other categories. These terms include common abbreviations such as `info' (information) and `id' (identification) and others that might not be so straightforward, like the method name \texttt{ran\_check}, where `ran' expands to the range or the method name \texttt{tz\_NYC}, where the two terms expand to timezone and New York City, respectively.
\end{itemize}

As part of our analysis of abbreviations and acronyms, we examined the extent to which these shortened terms are defined. For each identified abbreviation or acronym, we manually analyzed the notebook to locate the presence of its expansion. We envision that by knowing the typical locations of expansions, tools can be more accurate and efficient in expanding abbreviations, thereby improving developer comprehension. The distribution of locations is as follows:
\begin{itemize}
    \item \textbf{Markdown Cell} - These are formatted text sections in Jupyter notebooks that use Markdown for documentation. These cells contain definitions for 82 abbreviations and acronyms. For example, the markdown cell above the code cell containing the method \texttt{compute\_series\_cy}  has the text ``\textit{Now let's implement a Cython version},'' which provides the expansion for `cy' as Cython.  
    
    \item \textbf{Code Comment} - Our analysis identified 31 instances of abbreviation and acronym expansions in code comments, occurring either as inline comments or docstring comments. For example, the expansion of `lr' in the method name \texttt{lr\_schedule} can be inferred from the inline comment ``\textit{\# Define a learning rate schedule}.'' Unlike content in markdown cells, these comments can provide a more immediate definition of the shortened terms. 
    
    \item \textbf{Code Context} - Another technique to infer the expansion of abbreviations and acronyms is examining surrounding code elements. This inference can be achieved using contextual clues embedded within the method implementation, such as parameter and variable names and their assignments, method calls, and arguments passed in these calls. For example, consider the partial code snippet in Listing \ref{lst:rq3_code01}. The expansion of `cm' in the method name \texttt{plot\_cm} can be inferred through multiple contextual elements: (1) the default value of the \texttt{title} parameter (i.e., `Confusion Matrix') and (2) the name of the method call \texttt{confusion\_matrix}. We identified 23 expansions using this technique.

    \item \textbf{External Sources} - We encountered 94 instances where acronyms and abbreviations could not be expanded solely on the notebook's content. These shortened terms generally fell into two categories: common computing terms, such as `json' (JavaScript Object Notation), or domain-specific terms, such as `slam' (simultaneous localization and mapping). While common acronyms and abbreviations are less challenging to expand, domain-specific terms present a greater challenge, requiring specialized technical documentation or domain-specific resources. This pattern reflects an assumption that notebook readers are domain experts who would recognize these shortened terms. However, this assumption creates potential barriers for novices, interdisciplinary researchers, or maintainers from different backgrounds. For this study, we utilized internet search engines to identify expansions for the domain-specific terms.
\end{itemize}

\begin{lstlisting}[style=pythonstyle, caption={Inferring acronym expansion through surrounding code.}, label={lst:rq3_code01}]
def plot_cm(y_true, y_pred, title='Confusion Matrix', cmap=plt.cm.Blues):
    sns.set_style('white')
    cm = confusion_matrix(y_test, y_pred)
    # Additional implementation...
\end{lstlisting}

\begin{tcolorbox}[top=0.5pt,bottom=0.5pt,left=1pt,right=1pt]
\textbf{RQ3 Summary.}
Approximately 30\% of method names in notebooks contain abbreviations and/or acronyms in their names. These shortened terms fall into categories such as mathematics and statistics, image processing, programming, and artificial intelligence. While expansions can be found in markdown cells, code comments, and surrounding code elements, most abbreviations and acronyms require specialized knowledge to interpret, indicating that practitioners often assume readers possess domain expertise.
\end{tcolorbox}

\section{Discussion}
\label{Section:discussion}
The study reveals the unique characteristics of method names in Jupyter Notebooks and their impacts on comprehensibility. 
This section contrasts our results with those from related studies and discusses the implications of our findings.

\subsection{Prior Work Comparison}
As reported in Section \ref{Section:related}, this study is the first to examine the structure of method names within Jupyter Notebooks. As part of our discussion, we compare our findings against related research on naming in traditional software development.

\subsubsection{\textbf{Name Conciseness}}
Our findings from RQ1 indicate that method names in Jupyter Notebooks are generally short, with the median number of words in a name being two. While shorter names can result in more concise code, previous research on identifier naming suggests that shorter identifiers are less effective for code comprehension when compared to longer, more descriptive compound names \cite{Schankin2018,Hofmeister2017,Lawrie2007}. Additionally, a survey of 1,152 professional developers conducted by Alsuhaibani et al. \cite{Alsuhaibani2021} shows that developers prefer method names consisting of a median of five words. %three to seven words, with a median of five words.

\subsubsection{\textbf{Grammar Patterns}}
To compare the grammar patterns identified in RQ2, we analyze the dataset produced by Newman et al. \cite{NewmanCoRR2020}, which contains human-annotated grammar patterns from 267 method names across Java, C\#, C++, and C open-source systems. While Newman et al. identified 96 unique grammar patterns from their 267 methods, our analysis yielded 68 unique patterns from a larger sample of 691 methods. Our dataset shows higher concentration in top patterns, while their distribution is more evenly spread across various pattern types. These differences in pattern diversity, despite our larger sample size, suggest varying preferences in naming across different programming contexts. Below, we present some notable differences and similarities:% between these datasets.

\begin{itemize}
    \item \textbf{\textit{Most Frequent Patterns}:} Their dataset shows \textbf{V NM N} as the dominant pattern (17.23\%), while ours is \textbf{V-N}  (19.54\%). 
    
    \item \textbf{\textit{Pattern Complexity}:} There are more varying and complex patterns with longer chains of parts of speech in their dataset (e.g., \textbf{N V P V VM DT NPL}). In contrast, our patterns typically maintain simpler structures.
    
    \item \textbf{\textit{Preamble Usage}:} A notable difference is the occurrence of the preamble (\textbf{PRE}) tag in their dataset (e.g., \textbf{PRE V NM N}), of which we have none.
    
    \item \textbf{\textit{Single Noun Pattern}:} Another difference is that single noun (\textbf{N}) patterns are much more common in our dataset (12.74\%) than theirs (1.12\%).
    
    \item \textbf{\textit{Noun-Verb Patterns}:} While both datasets contain patterns of a verb following a noun (\textbf{N V}), the implementation differs: (1) this pattern in our dataset occurs at 2.89\%, while in their dataset it is at 1.87\% and (2) their dataset includes more complex variations of this pattern (e.g., \textbf{NM NM N V}, \textbf{PRE NM N V}).

    \item \textbf{\textit{Verb Modifier Usage}:} Both datasets have instances of the verb modifier tag (\textbf{VM}), but in our method names, it appears at the beginning, whereas in their dataset, it appears within the name (e.g., \textbf{V VM N}).
\end{itemize}

\subsubsection{\textbf{Abbreviation \& Acronym Expansion}}
Our findings on abbreviation and acronym expansion share similarities with a study by Newman et al. \cite{NewmanICSME2019} on abbreviations in Java, C, and C++ systems. They identified multiple sources for locating abbreviation expansions: source code, comments, language documentation, project documentation, and dictionaries. While the general categories of sources overlap, the notebook environment differs significantly from traditional software systems, where documentation is spread across multiple files (API docs, README files, etc.). In contrast, notebooks are self-contained and use markdown cells for documentation, meaning that abbreviations and their meanings are usually understood within the context of a single notebook, unlike in separate documents. Further, Newman et al. rely on computer science and English dictionaries to expand abbreviations. While these sources will be useful for notebook environments, they are unlikely to include domain-specific abbreviations and acronyms, highlighting a unique challenge in expanding shortened terms within the scientific computing context.

\subsection{Implications}
Our findings have several important implications:

\subsubsection{\textbf{Extending the Catalog of Linguistic Antipatterns}}
While analyzing grammar patterns and identifying deviations, we also encountered various naming violations referred to as linguistic antipatterns \cite{Arnaoudova2016}. %\rev{Linguistic antipatterns are inconsistencies between the name of an identifier and its function within the code, potentially misleading and impairing program understanding.} 
Although this was not the primary focus of our study, we share some of our observations.  For example, the method \texttt{get\_prediction} lacks a return statement, which classifies it as a linguistic antipattern known as ``get method does not return.'' Likewise, the method \texttt{get\_target} returns a list (i.e., \texttt{return list(target\_words)}), yet the term `target' is singular, despite the method returning a collection; this represents the ``expecting but not getting a single instance'' antipattern. A more appropriate name for this method would be `get\_target\_words.' In addition to these established linguistic antipatterns, we also identified potentially new ones, expanding upon the original catalog. Specifically: 
\begin{itemize}
    \item Consider the method \texttt{play\_separated\_track} that returns an array. Although the name follows the standard method grammar pattern (i.e., verb phrase), the verb `play' suggests an action of audio playback rather than a data retrieval operation. This creates a linguistic antipattern we classify as `\textit{`Non-retrieval Action Verb Returns Data}.'' This type of antipattern obscures the method's data-returning nature, potentially creating misleading expectations.
    \item Consider the method \texttt{perform\_testing} that evaluates a neural network on a test dataset. Although the name adheres to the verb phrase pattern, the term `testing' creates ambiguity in the machine learning context as the term is overloaded between software testing and model evaluation. In contrast, the method \texttt{test\_train} tests the correctness of the training process as part of software testing. However, its name also has similar ambiguity due to combining `test' and `train.' This represents a linguistic antipattern we classify as ``\textit{Term Has Competing Meanings},''  and depending on the domain context, could lead to significant confusion.
\end{itemize}

\subsubsection{\textbf{The Need for Code Readability Models for Scientific Computing}} 
Existing code readability models have been developed primarily for general-purpose software development contexts, with most of them trained on Java code \cite{Buse2008,Posnett2011,Scalabrino2018,Qing2022}. While these models are useful, they may not adequately capture the unique characteristics of scientific code, especially in notebook environments. Notebook code is typically less complex, but more entangled than the code found in classical programs \cite{grotov2022large}. In addition, the model's features, such as line length, indentation, and the presence of comments, while also applicable to code in notebooks, are not sufficient. Jupyter Notebooks contain not only code but also markdown and output cells, which are not considered in these models, which typically evaluate code in isolation, focusing on line-by-line readability. This technique might not be effective in a notebook environment due to the contextual information provided by the surrounding markdown cells \cite{Rule2018}. For example, our RQ 3 findings show that markdown cells can be a source for expanding abbreviations and acronyms found in method names.
Further, the high prevalence of domain-specific terminology, mathematical notation, and other specialized abbreviations and acronyms in notebooks becomes challenging for models that rely on dictionary terms unless they are fine-tuned with relevant datasets that include such terms. For example, in the method name \texttt{apply\_sobel}, the term `Sobel' is neither an acronym nor abbreviation but refers to the ``Sobel operator,'' a technique used in image processing and computer vision. Likewise, the digit `2' in the name \texttt{custom\_f2} refers to a specific variant of the F-score metric in machine learning rather than serving as a counter or index. Such examples highlight the need for contextual understanding when interpreting domain-specific terminology in technical fields.

\subsubsection{\textbf{Educational Resources for Improving Scientific Code Readability}} Our findings reveal numerous deviations from traditional naming practices in scientific code, highlighting a gap between software engineering and scientific programming best practices. This presents an opportunity to develop targeted educational resources for scientific programmers, who often lack formal computer science training and primarily use notebooks for rapid prototyping and interactive data exploration \cite{Perkel2018,Santana2024}. While our findings emphasize the need for educating scientific programmers on the need and types of high-quality naming patterns, the training should also extend to other areas of code quality that typically impact notebooks \cite{grotov2022large}. Further, even though our study focused on method naming patterns, we observed that markdown cells play an important role in code comprehension. These cells often contain essential context for domain and scientific concepts and implementation choices, showing the importance of educational resources that focus on both code quality and effective documentation practices.

\section{Threats To Validity}
\label{Section:threats}
In this exploratory study, we analyzed a statistically significant random sample of notebooks taken from a published dataset consisting of 847,881 notebooks \cite{grotov2022large} that have been utilized in prior research \cite{Montandon2023, Grotov2023}. However, this introduces a threat to the generalizability of our results, as the dataset is restricted to Python-based Jupyter Notebooks from GitHub public repositories, which may not represent naming practices in private repositories, other notebook environments,% (e.g., Google Colab), 
or notebooks written in other programming languages. Additionally, while our random sampling approach ensures statistical significance, it may have inadvertently excluded notebooks that could provide valuable insights into naming practices. %Further, the source dataset's lack of author information prevented us from contacting authors to understand their naming decisions.% or examining broader project context.

Our manual annotation process introduces potential threats through subjective interpretations in assigning part-of-speech tags and identifying abbreviations and acronyms, as well as possible bias during disagreement resolution discussions. Since annotations were performed by the authors rather than independent domain experts, there is also a risk of confirmation bias. To mitigate these threats, we (1) required each method name to be independently annotated by different annotators, (2) cross-validated all annotations, (3) provided annotators with access to each notebook, and (4) established annotation guidelines and examples. However, it is possible that multiple valid grammatical interpretations of the same method name may exist, which can arise when the programmer's intended interpretation differs from the name's structural composition. 
Lastly, our research scope may not capture all aspects of method naming practices. The grammar pattern analysis primarily focused on syntactic structure rather than complete semantics and may overlook some domain-specific conventions. Likewise, our classification of abbreviations and acronyms might not cover all areas of scientific computing.

\section{Conclusion \& Future Work}
\label{Section:conclusion}
Methods are fundamental programming components that require appropriate naming to convey their purpose. This exploratory study provides the first systematic analysis of method naming practices in Jupyter Notebooks, where we analyze 691 method names across 384 notebooks. 

Our key findings reveal that while scientific programmers generally follow Python's naming style conventions, they often deviate from established method naming best practices. Only 55.57\% of names begin with verbs, and names tend to be shorter than in traditional software development, with 48.77\% containing just two terms. Further, 30.39\% of method names include abbreviations or acronyms, related to areas such as mathematics or statistics, image processing, and artificial intelligence, among others. We envision our findings facilitating the improvement of code quality tools for the scientific domain, developing educational resources tailored to scientific programmers, and furthering research in this area. 

Our future work will focus on validating these results through user studies with notebook practitioners. Through surveys and interviews, we aim to understand practitioners' rationale behind naming choices and explore how naming patterns impact code comprehension. %Finally, our annotated dataset is available at \cite{ArtifactPackage}, enabling replication and extension of this work.

\bibliographystyle{ieeetr}
\bibliography{main}

\begin{thebibliography}{10}

\bibitem{mcconnell2004code}
S.~McConnell, {\em Code Complete}.
\newblock Developer Best Practices, Pearson Education, 2004.

\bibitem{martin2009clean}
R.~Martin, {\em Clean Code: A Handbook of Agile Software Craftsmanship}.
\newblock Robert C. Martin series, Prentice Hall, 2009.

\bibitem{osherove2013art}
R.~Osherove, {\em The Art of Unit Testing: with examples in C\#}.
\newblock Manning, 2013.

\bibitem{Host2009}
E.~W. H{\o}st and B.~M. {\O}stvold, ``Debugging method names,'' in {\em ECOOP 2009 -- Object-Oriented Programming}, (Berlin, Heidelberg), pp.~294--317, Springer Berlin Heidelberg, 2009.

\bibitem{Host2007}
E.~W. H{\o}st and B.~M. {\O}stvold, ``The programmer's lexicon, volume i: The verbs,'' in {\em Seventh IEEE International Working Conference on Source Code Analysis and Manipulation (SCAM 2007)}, pp.~193--202, 2007.

\bibitem{Feitelson2023}
D.~G. Feitelson, ``From code complexity metrics to program comprehension,'' {\em Commun. ACM}, vol.~66, p.~52–61, Apr. 2023.

\bibitem{LawrieICPC2006}
D.~Lawrie, C.~Morrell, H.~Feild, and D.~Binkley, ``What's in a name? a study of identifiers,'' in {\em 14th IEEE International Conference on Program Comprehension (ICPC'06)}, pp.~3--12, 2006.

\bibitem{ButlerCSMR2010}
S.~Butler, M.~Wermelinger, Y.~Yu, and H.~Sharp, ``Exploring the influence of identifier names on code quality: An empirical study,'' in {\em 2010 14th European Conference on Software Maintenance and Reengineering}, pp.~156--165, 2010.

\bibitem{Li2020Renamings}
G.~Li, H.~Liu, and A.~S. Nyamawe, ``A survey on renamings of software entities,'' {\em ACM Comput. Surv.}, vol.~53, Apr. 2020.

\bibitem{Alsuhaibani2021}
R.~Alsuhaibani, C.~Newman, M.~Decker, M.~Collard, and J.~Maletic, ``On the naming of methods: A survey of professional developers,'' in {\em 2021 IEEE/ACM 43rd International Conference on Software Engineering (ICSE)}, pp.~587--599, 2021.

\bibitem{gresta2023naming}
R.~Gresta, V.~Durelli, and E.~Cirilo, ``Naming practices in object-oriented programming: An empirical study,'' {\em Journal of Software Engineering Research and Development}, vol.~11, no.~1, pp.~5--1, 2023.

\bibitem{binkley-emse13}
D.~W. Binkley, M.~Davis, D.~J. Lawrie, J.~I. Maletic, C.~Morrell, and B.~Sharif, ``The impact of identifier style on effort and comprehension,'' {\em Empir. Softw. Eng.}, vol.~18, no.~2, pp.~219--276, 2013.

\bibitem{Hofmeister2017}
J.~Hofmeister, J.~Siegmund, and D.~V. Holt, ``Shorter identifier names take longer to comprehend,'' in {\em 2017 IEEE 24th International Conference on Software Analysis, Evolution and Reengineering (SANER)}, pp.~217--227, 2017.

\bibitem{Rule2018}
A.~Rule, A.~Tabard, and J.~D. Hollan, ``Exploration and explanation in computational notebooks,'' in {\em Proceedings of the 2018 CHI Conference on Human Factors in Computing Systems}, CHI '18, (New York, NY, USA), p.~1–12, Association for Computing Machinery, 2018.

\bibitem{Perkel2018}
J.~M. Perkel, ``Why jupyter is data scientists’ computational notebook of choice,'' {\em Nature}, vol.~563, p.~145–146, Oct. 2018.

\bibitem{Kery2018}
M.~B. Kery, M.~Radensky, M.~Arya, B.~E. John, and B.~A. Myers, ``The story in the notebook: Exploratory data science using a literate programming tool,'' in {\em Proceedings of the 2018 CHI Conference on Human Factors in Computing Systems}, CHI '18, (New York, NY, USA), p.~1–11, Association for Computing Machinery, 2018.

\bibitem{Santana2024}
T.~L. De~Santana, P.~A. D. M.~S. Neto, E.~S. De~Almeida, and I.~Ahmed, ``Bug analysis in jupyter notebook projects: An empirical study,'' {\em ACM Trans. Softw. Eng. Methodol.}, vol.~33, Apr. 2024.

\bibitem{Beg2021}
M.~Beg, J.~Taka, T.~Kluyver, A.~Konovalov, M.~Ragan-Kelley, N.~M. Thiéry, and H.~Fangohr, ``Using jupyter for reproducible scientific workflows,'' {\em Computing in Science \& Engineering}, vol.~23, no.~2, pp.~36--46, 2021.

\bibitem{Samuel2024}
S.~Samuel and D.~Mietchen, ``Computational reproducibility of jupyter notebooks from biomedical publications,'' {\em Gigascience}, vol.~13, Jan. 2024.

\bibitem{Wang2020}
J.~Wang, L.~Li, and A.~Zeller, ``Better code, better sharing: on the need of analyzing jupyter notebooks,'' in {\em Proceedings of the ACM/IEEE 42nd International Conference on Software Engineering: New Ideas and Emerging Results}, ICSE-NIER '20, (New York, NY, USA), p.~53–56, Association for Computing Machinery, 2020.

\bibitem{PimentelMSR2019}
J.~F. Pimentel, L.~Murta, V.~Braganholo, and J.~Freire, ``A large-scale study about quality and reproducibility of jupyter notebooks,'' in {\em 2019 IEEE/ACM 16th International Conference on Mining Software Repositories (MSR)}, pp.~507--517, 2019.

\bibitem{grotov2022large}
K.~Grotov, S.~Titov, V.~Sotnikov, Y.~Golubev, and T.~Bryksin, ``A large-scale comparison of python code in jupyter notebooks and scripts,'' in {\em Proceedings of the 19th International Conference on Mining Software Repositories}, MSR '22, (New York, NY, USA), p.~353–364, Association for Computing Machinery, 2022.

\bibitem{Adams2023}
K.~Adams, A.~Vilkomir, and M.~Hills, ``A comparison of machine learning code quality in python scripts and jupyter notebooks,'' {\em J. Comput. Sci. Coll.}, vol.~39, p.~96–108, Nov. 2023.

\bibitem{Butler2009}
S.~Butler, M.~Wermelinger, Y.~Yu, and H.~Sharp, ``Relating identifier naming flaws and code quality: An empirical study,'' in {\em 2009 16th Working Conference on Reverse Engineering}, pp.~31--35, 2009.

\bibitem{Arnaoudova2016}
V.~Arnaoudova, M.~Di~Penta, and G.~Antoniol, ``Linguistic antipatterns: what they are and how developers perceive them,'' {\em Empirical Software Engineering}, vol.~21, pp.~104--158, Feb 2016.

\bibitem{Allamanis2015}
M.~Allamanis, E.~T. Barr, C.~Bird, and C.~Sutton, ``Suggesting accurate method and class names,'' in {\em Proceedings of the 2015 10th Joint Meeting on Foundations of Software Engineering}, ESEC/FSE 2015, (New York, NY, USA), p.~38–49, Association for Computing Machinery, 2015.

\bibitem{Binkley2009}
D.~Binkley, M.~Davis, D.~Lawrie, and C.~Morrell, ``To camelcase or under\_score,'' in {\em 2009 IEEE 17th International Conference on Program Comprehension}, pp.~158--167, 2009.

\bibitem{Quaranta2022}
L.~Quaranta, F.~Calefato, and F.~Lanubile, ``Eliciting best practices for collaboration with computational notebooks,'' {\em Proc. ACM Hum.-Comput. Interact.}, vol.~6, Apr. 2022.

\bibitem{NewmanCoRR2020}
C.~D. Newman, R.~S. AlSuhaibani, M.~J. Decker, A.~Peruma, D.~Kaushik, M.~W. Mkaouer, and E.~Hill, ``On the generation, structure, and semantics of grammar patterns in source code identifiers,'' {\em Journal of Systems and Software}, vol.~170, p.~110740, 2020.

\bibitem{PerumaICPC2021}
A.~Peruma, E.~Hu, J.~Chen, E.~A. AlOmar, M.~W. Mkaouer, and C.~D. Newman, ``Using grammar patterns to interpret test method name evolution,'' in {\em 2021 IEEE/ACM 29th International Conference on Program Comprehension (ICPC)}, pp.~335--346, 2021.

\bibitem{Arnaoudova2014}
V.~Arnaoudova, L.~M. Eshkevari, M.~D. Penta, R.~Oliveto, G.~Antoniol, and Y.-G. Guéhéneuc, ``Repent: Analyzing the nature of identifier renamings,'' {\em IEEE Transactions on Software Engineering}, vol.~40, no.~5, pp.~502--532, 2014.

\bibitem{Butler2011}
S.~Butler, M.~Wermelinger, Y.~Yu, and H.~Sharp, ``Mining java class naming conventions,'' in {\em 2011 27th IEEE International Conference on Software Maintenance (ICSM)}, pp.~93--102, 2011.

\bibitem{Caprile1999}
C.~Caprile and P.~Tonella, ``Nomen est omen: analyzing the language of function identifiers,'' in {\em Sixth Working Conference on Reverse Engineering (Cat. No.PR00303)}, pp.~112--122, 1999.

\bibitem{Host2009PhraseBook}
E.~W. H{\o}st and B.~M. {\O}stvold, ``The java programmer's phrase book,'' in {\em Software Language Engineering} (D.~Ga{\v{s}}evi{\'{c}}, R.~L{\"a}mmel, and E.~Van~Wyk, eds.), (Berlin, Heidelberg), pp.~322--341, Springer Berlin Heidelberg, 2009.

\bibitem{NewmanICSME2019}
C.~D. Newman, M.~J. Decker, R.~S. Alsuhaibani, A.~Peruma, D.~Kaushik, and E.~Hill, ``An empirical study of abbreviations and expansions in software artifacts,'' in {\em 2019 IEEE International Conference on Software Maintenance and Evolution (ICSME)}, pp.~269--279, 2019.

\bibitem{Montandon2023}
J.~E. Montandon, L.~L. Silva, C.~Politowski, G.~E. Boussaidi, and M.~T. Valente, ``Unboxing default argument breaking changes in scikit learn,'' in {\em 2023 IEEE 23rd International Working Conference on Source Code Analysis and Manipulation (SCAM)}, pp.~209--219, 2023.

\bibitem{Grotov2023}
K.~Grotov, S.~Titov, A.~Suhinin, Y.~Golubev, and T.~Bryksin, ``Optimizing duplicate size thresholds in ides,'' in {\em 2023 IEEE/ACM 20th International Conference on Mining Software Repositories (MSR)}, pp.~470--471, 2023.

\bibitem{Olney2016}
W.~Olney, E.~Hill, C.~Thurber, and B.~Lemma, ``Part of speech tagging java method names,'' in {\em 2016 IEEE International Conference on Software Maintenance and Evolution (ICSME)}, pp.~483--487, 2016.

\bibitem{Newman2022}
C.~D. Newman, M.~J. Decker, R.~S. Alsuhaibani, A.~Peruma, M.~W. Mkaouer, S.~Mohapatra, T.~Vishnoi, M.~Zampieri, T.~J. Sheldon, and E.~Hill, ``An ensemble approach for annotating source code identifiers with part-of-speech tags,'' {\em IEEE Transactions on Software Engineering}, vol.~48, no.~9, pp.~3506--3522, 2022.

\bibitem{SCANLCatalogue}
SCANL, ``Identifier name structure catalogue.'' \url{https://github.com/SCANL/identifier_name_structure_catalogue}, 2022.

\bibitem{fleiss1971measuring}
J.~L. Fleiss, ``Measuring nominal scale agreement among many raters.,'' {\em Psychological bulletin}, vol.~76, no.~5, p.~378, 1971.

\bibitem{kappaInterpret}
J.~R. Landis and G.~G. Koch, ``The measurement of observer agreement for categorical data,'' {\em Biometrics}, vol.~33, no.~1, pp.~159--174, 1977.

\bibitem{ArtifactPackage}
``Artifact package.'' \url{https://doi.org/10.5281/zenodo.14800298}.

\bibitem{Binkley2011}
D.~Binkley, M.~Hearn, and D.~Lawrie, ``Improving identifier informativeness using part of speech information,'' in {\em Proceedings of the 8th Working Conference on Mining Software Repositories}, MSR '11, (New York, NY, USA), p.~203–206, Association for Computing Machinery, 2011.

\bibitem{Host2011}
E.~W. H{\o}st and B.~M. {\O}stvold, ``Canonical method names for java,'' in {\em Software Language Engineering} (B.~Malloy, S.~Staab, and M.~van~den Brand, eds.), (Berlin, Heidelberg), pp.~226--245, Springer Berlin Heidelberg, 2011.

\bibitem{Schankin2018}
A.~Schankin, A.~Berger, D.~V. Holt, J.~C. Hofmeister, T.~Riedel, and M.~Beigl, ``Descriptive compound identifier names improve source code comprehension,'' in {\em Proceedings of the 26th Conference on Program Comprehension}, ICPC '18, (New York, NY, USA), p.~31–40, Association for Computing Machinery, 2018.

\bibitem{Lawrie2007}
D.~Lawrie, C.~Morrell, H.~Feild, and D.~Binkley, ``Effective identifier names for comprehension and memory,'' {\em Innovations in Systems and Software Engineering}, vol.~3, pp.~303--318, Dec 2007.

\bibitem{Buse2008}
R.~P. Buse and W.~R. Weimer, ``A metric for software readability,'' in {\em Proceedings of the 2008 International Symposium on Software Testing and Analysis}, ISSTA '08, (New York, NY, USA), p.~121–130, Association for Computing Machinery, 2008.

\bibitem{Posnett2011}
D.~Posnett, A.~Hindle, and P.~Devanbu, ``A simpler model of software readability,'' in {\em Proceedings of the 8th Working Conference on Mining Software Repositories}, MSR '11, (New York, NY, USA), p.~73–82, Association for Computing Machinery, 2011.

\bibitem{Scalabrino2018}
S.~Scalabrino, M.~Linares-Vásquez, R.~Oliveto, and D.~Poshyvanyk, ``A comprehensive model for code readability,'' {\em Journal of Software: Evolution and Process}, vol.~30, no.~6, p.~e1958, 2018.
\newblock e1958 smr.1958.

\bibitem{Qing2022}
Q.~Mi, Y.~Hao, L.~Ou, and W.~Ma, ``Towards using visual, semantic and structural features to improve code readability classification,'' {\em Journal of Systems and Software}, vol.~193, p.~111454, 2022.

\end{thebibliography}
\end{document}